\documentclass{article}
\usepackage{sprocl2e}
\usepackage{cite}

\def\prd#1#2#3{{\sl Phys.~Rev.~}{\bf D#1} (#2) #3}
\def\prl#1#2#3{{\sl Phys.~Rev.~Lett.~}{\bf #1} (#2) #3}

\def\be{\begin{equation}}
\def\ee{\end{equation}}
\def\bea{\begin{eqnarray}}
\def\eea{\end{eqnarray}}

\newenvironment{Eqnarray}%
         {\arraycolsep 0.14em\begin{eqnarray}}{\end{eqnarray}}
\def\beqa{\begin{Eqnarray}}
\def\eeqa{\end{Eqnarray}}
\def\ifmath#1{\relax\ifmmode #1\else $#1$\fi}
\def\ls#1{\ifmath{_{\lower1.5pt\hbox{$\scriptstyle #1$}}}}
\def\hsm{h\ls{\rm SM}}
\renewcommand{\thefootnote}{\arabic{footnote}}
\long\def\symbolfootnote[#1]#2{\begingroup%
\def\thefootnote{\fnsymbol{footnote}}\footnote[#1]{#2}\endgroup} 

\begin{document}
\vbox{  \large
\begin{flushright}
SCIPP 04/13 \\
August, 2004 \\
hep--ph/0409008\\
\end{flushright}
\vskip1.5cm
\begin{center}
{\LARGE\bf
Higgs Theory---A Brief Overview} \\[1cm]

{\Large Howard E. Haber}\\[5pt]{\it Santa Cruz Institute for Particle
Physics,  \\ University of California, Santa Cruz, CA 95064, U.S.A.}\\

\vskip2cm
\thispagestyle{empty}

{\bf Abstract}\\[1pc]

\begin{minipage}{10cm}
A brief overview is given of the theory of Higgs bosons 
and electroweak symmetry breaking that is
relevant for the Higgs physics program at the Linear Collider. 
\end{minipage}  \\
\vskip4cm
Invited talk at the \\
International Conference on Linear Colliders LCWS-04 \\
19--23 April 2004, Paris, France \\
\end{center}
}
\vfill
\clearpage
\setcounter{page}{1}

\title{HIGGS THEORY---A BRIEF OVERVIEW~\symbolfootnote[1]{This 
work was supported in part by the U.S. DOE
grant no.~DE-FG03-92ER40689.
}
\\}

\author{HOWARD E.~HABER}

\address{Santa Cruz Institute for Particle Physics\\
University of California, Santa Cruz, CA 95064 USA}


\maketitle\abstracts{
A brief overview is given of the theory of Higgs bosons 
and electroweak symmetry breaking that is
relevant for the Higgs physics program at the Linear Collider. 
}

\section{Introduction}

One of
the highest priorities of particle physics today is the discovery of
the dynamics responsible for electroweak symmetry breaking (EWSB).
In the Standard Model (SM), this dynamics is achieved by the
self-interactions of a complex scalar doublet of fields.  This approach
predicts the existence of one physical
elementary scalar---the Higgs boson \cite{hhg}.  
It has been argued that if
the mass of the SM Higgs boson ($\hsm$) lies between about 
130 and 200~GeV~\cite{riesselmann},
then the SM can be valid at energy scales all the way up
to the Planck scale.  

However, it is difficult to imagine a fundamental theory of elementary
particles and their interactions with no explanation for the origin of
the large hierarchy of mass scales from $m_Z$ to the Planck mass.
Thus, most theorists expect new physics beyond the SM at the TeV-scale
to emerge and provide a ``natural'' explanation of the connection
between these two disparate mass scales.  To fully probe the nature of
EWSB and the associated new TeV-scale physics, one must conduct
experiments at the LHC and the LC.

\section{Can A Light Higgs Boson Be Avoided?}

Based on the most recent fits to electroweak data, the LEP Electroweak
Working Group \cite{lepewwg} 
concludes that ``the Standard Model is able to describe
nearly all the LEP measurements rather well,'' with no compelling need
for physics beyond the SM.  This analysis yields a prediction for the
Higgs mass: $m_{\hsm}=114^{+69}_{-45}$~GeV or a one-sided 95\% CL upper
limit of $m_{\hsm}<260$~GeV.  These results have definite consequences for
the anticipated Higgs studies at the LHC and LC, so it is natural to
ask whether these conclusions can be avoided.  

The probability of the goodness of the SM electroweak
fit based only on high $Q^2$
data is 26\% (this figure is reduced if the NuTeV results are
included).  
Taking the SM electroweak fit at face value,
one can use the data to constrain theories of
new physics (whose low-energy effective theory closely
approximates the SM).  An example of such a procedure
employs the $S$ and $T$ parameters of Peskin and Takeuchi \cite{stu}, 
under the assumption that the main effects of the new physics enter
through the modification of the $W$ and $Z$ boson self-energies.
The precision electroweak
data impose strong constraints on any new physics beyond the
SM.  For example, if the Higgs mass is significantly larger than the upper
bound quoted above, new physics beyond the SM must
contribute positively to $T$ (and perhaps negatively to $S$) in order
to be consistent with the precision electroweak fits.

\section{The Nature Of EWSB Dynamics}

In addition to scalar dynamics of the SM, there have been many
theories proposed in the literature to explain the mechanism of EWSB.
Some theories employ weakly-coupled scalar dynamics, while others employ
strongly-coupled dynamics of a new sector of particles.  
The motivation of nearly all proposed theories of
EWSB beyond the SM is to address the theoretical problems of
naturalness and hierarchy.  The four main theoretical approaches are
as follows:

\textbf{Higgs bosons of low-energy supersymmetry} \cite{Carena:2002es}.
As in the SM, these models employ weakly-coupled scalar dynamics, and
all Higgs scalars are elementary particles.  Supersymmetry eliminates
all quadratic sensitivity to the Planck scale, at the price of
TeV-scale supersymmetry breaking whose fundamental origin is unknown.

\textbf{Little Higgs models} \cite{littlehiggs}.
The light Higgs bosons of these models are nearly indistinguishable
from the elementary scalars of weakly-coupled EWSB theories.  However,
new physics must enter near the TeV scale to cancel out one-loop quadratic
sensitivity of the theory to the ultraviolet scale.
These theories have an implicit cutoff of about 10~TeV, above which
one would need to find their ultraviolet completions.  

\textbf{Extra-dimensional theories of EWSB} \cite{extradim}.
Such approaches lead to new models of EWSB dynamics, including the
so-called ``Higgsless'' models \cite{higgsless}
in which there is no light Higgs scalar
in the spectrum.  Such models also require an ultraviolet completion
at a scale characterized by the inverse radius of the extra dimension.

\textbf{Strongly-coupled EWSB sectors} \cite{strong}.
Models of this type include technicolor models, composite Higgs models
of various kinds, top-quark condensate models, \textit{etc.}

New physics beyond the SM can be of two
types---decoupling or non-decoupling.  
The virtual effects of ``decoupling'' physics beyond the SM
typically scale as
$m_Z^2/M^2$, where $M$ is a scale characteristic of the new physics.
Examples of this type include ``low-energy'' supersymmetric theories
with soft-supersymmetry-breaking masses of $\mathcal{O}(M)$.  In
contrast, some of the virtual effects of ``non-decoupling'' physics 
do not vanish as the characteristic scale $M\to\infty$.  A theory
with a fourth generation fermion and technicolor models are examples
of this type.  Clearly, the success of the SM electroweak fit places
stronger restrictions on non-decoupling new physics.  Nevertheless,
some interesting constraints on decoupling physics can also be
obtained.  For example, even in theories of new physics that exhibit
decoupling, the scale $M$ must be somewhat separated from the scale $m_Z$ 
(to avoid conflict with the SM electroweak fit).  This leads to a
tension with the requirements of naturalness, which has been called
the ``little hierarchy problem'' \cite{lhp} in the literature.

\section{Approaching The Decoupling Limit}

Many models of EWSB yield a lightest Higgs boson whose properties are
nearly identical to those of the SM Higgs boson (the
so-called {\textit{decoupling limit}} \cite{decoupling}).  
Thus, to probe the physics of
EWSB, one must either detect deviations of the lightest CP-even Higgs
boson from the decoupling limit and/or directly observe the additional
degrees of freedom associated with the EWSB sector.  The latter is
expected to be connected with the TeV-scale physics responsible for a
natural explanation of the electroweak scale.  Examples include:
non-minimal Higgs states (additional CP-even neutral scalars, CP-odd
scalars and charged scalars), supersymmetric particles, new gauge
bosons, vector-like fermions, Kaluza-Klein excitations and radions.

As an example consider the Higgs sector of the minimal supersymmetric
extension of the Standard Model (MSSM), which 
in the decoupling limit contains a CP-even Higgs boson, $h$, with
properties nearly identical to the SM Higgs boson.  The decoupling
limit is achieved in the limit of $m_A\gg m_Z$, where $m_A$ is the
mass of the CP-odd scalar of the model.  To illustrate these
statements, we exhibit the following ratio of MSSM Higgs 
couplings \cite{chlm}~\footnote{Eqs.~(\ref{ei}) and (\ref{eii}) include
the leading $\tan\beta$-enhanced radiative corrections, where
$\tan\beta$ is the ratio of the two neutral Higgs vacuum expectation values.}
\beqa
&&   {g^2_{hVV}\over g^2_{\hsm VV}}  \simeq 
1-{c^2 m_Z^4\sin^2 4\beta\over 4m_A^4}\,, \qquad\quad
   {g^2_{htt}\over g^2_{\hsm tt}}  \simeq  1+{c m_Z^2\sin 4\beta
\cot\beta\over m_A^2}\,,  \label{ei} \\[5pt]
&&  \qquad\qquad\qquad
{g^2_{hbb}\over g^2_{\hsm bb}}  \simeq  1-{4c m_Z^2\cos 2\beta
\over m_A^2}\left[\sin^2\beta-{\Delta_b\over 1+\Delta_b}\right]\,,
\label{eii}
\eeqa
where $c\equiv 1+\mathcal{O}(g^2)$ and $\Delta_b\equiv
\tan\beta\times \mathcal{O}(g^2)$ [here, $g$ is a generic gauge or
Yukawa coupling].  The quantities $c$ and $\Delta_b$ depend on the
MSSM spectrum. The approach to decoupling is fastest for the $h$
couplings to vector boson pairs and slowest for the couplings to
down-type quarks.
More generally, deviations from the
decoupling limit implicitly contain information about the EWSB
sector and the associated TeV-scale dynamics. 

\section{Main Goals Of The Higgs Hunter}

In preparing for a program of Higgs physics at 
present day and future colliders, 
one must first determine the discovery
reach of the colliders (Tevatron, LHC, LC, $\ldots$) for the
$\hsm$ and any additional states of a non-minimal Higgs sector that
might exist.  If evidence for the latter are found, one must 
then establish the
number of such states in the low-energy spectrum.
Once a candidate scalar state is discovered, one should ask whether it
is a Higgs boson (and whether it could be \textit{the} SM Higgs
boson).  Evidence for a non-minimal Higgs sector could emerge if
deviations from SM Higgs behavior is observed or Higgs states beyond
the $\hsm$ are produced.  

The decoupling limit, in which the lightest Higgs state closely
resembles the $\hsm$, presents an especially difficult challenge for
interpreting the underlying EWSB dynamics, since numerous theoretical
approaches can yield a lightest Higgs boson that is nearly indistinguishable
from the $\hsm$.  Nevertheless,
small deviations from the SM encode the physics of
EWSB and new physics beyond the SM.  In this case, 
a program of precision Higgs measurements 
at the LC is essential.  One will need to 
accurately measure
the mass, width, branching ratios and couplings of the candidate Higgs
states.  One must check the spin and CP-quantum numbers (keeping an
eye out for potential CP-violating effects).  Ideally, one would like
to reconstruct the full Higgs potential and directly verify the nature of the
EWSB scalar dynamics.  

If nature chooses a path far from the decoupling limit, the
challenges are of a different nature.  One must first determine if there
there are any light scalar states that are associated with EWSB
dynamics.  It may be that the theory of EWSB is based on
an extended Higgs sector
far from the decoupling limit.  In this case, one expects numerous light
Higgs states accessible to both the LHC and LC.  Otherwise, 
one must identify the source of EWSB dynamics and any associated phenomena.
In all such cases, one expects to have many light
states with a rich phenomenology anticipated at the 
LHC and LC,\footnote{Here, one needs further theoretical
study to see if there are viable counter-examples (perhaps Higgsless
theories?) in which no new physics below, say, 1 TeV is present.}
and precision measurements will again be essential in order to
distinguish among different theoretical approaches and models.


\section{Conclusions}

Recent approaches to EWSB dynamics have led to many new ideas and
models. Theorists have an important role to play in the development of
strategies for studying EWSB at future colliders.  We must begin to
systematize the attendant phenomenology that appears in new
approaches, with an eye toward finding some universal features.  We
must also devise new phenomenological techniques for distinguishing
among the various new approaches.  The precision Higgs studies can be
employed in this regard, although it is important to identify where
conventional assumptions could fail.  Finally, it will be especially
fruitful to refine and extend the studies, initiated in \cite{lhclc}, 
of the complementarity and interplay
of the LHC and LC searches for EWSB dynamics.
These will surely be significant steps toward unlocking the secrets of
the TeV scale.

%

\end{document}